# THE MATHEMATICAL THEORY OF DIFFUSION IN SOLIDS: TIME DEPENDENT FIRST KIND BOUNDARY CONDITIONS

**Guglielmo Macrelli (*)**

(*) Isoclima SpA – R&D Dept. Via A.Volta 14, 35042 Este (PD) Italy

guglielmomacrelli@hotmail.com

**Abstract**

A new solution to the mono-dimensional diffusion equation for time-variable first kind boundary condition is presented where the time-variable function at the surface is derived proposing a surface saturation model. This solution may be helpful in the treatment of diffusion processes where the overall time of diffusion is comparable with the time taken by the surface of the solid body to saturate achieving a dynamical equilibrium between the diffusing elements supplied by the external source and the ones transferred internally through the diffusion kinetic mechanisms. Worked examples for constant diffusion coefficient are presented and discussed.

## I. Introduction

Diffusion in solids is a subject largely treated in the literature [1],[2],[3],[4],[5],[6]. At the same time, the mathematics of diffusion is a topic well covered in the literature [7],[8],[9],[10]. It is remarkable that significant progresses in the mathematical analysis of diffusion equation can be found in the treatment of heat conduction problems [7],[9]. Formally the diffusion equation is mathematically equal to the heat conduction equation where concentration is substituted by temperature and the diffusion coefficient is substituted by the thermal diffusivity. This similarity can be addressed to the similarity of the constitutive equations for the material flux (first Fick law) and for the heat flux (Fourier law). The purpose of this paper is to review the diffusion equation derived from balance equation and constitutive equation, and to analyze solution with time dependent first kind boundary condition. The rationale of this approach is to relax the limitation of time independent, constant boundary condition of first kind. This last approach, even though pretty popular in the literature, represents an approximation based on the assumption of instantaneous achievement at initial time of a fixed value of surface concentration and keeping of its constancy over time when the solid is exposed to the source of diffusing elements. This last assumption appears can be justified only when the time to reach surface equilibrium, let's call it saturation time ($\tau_{sat}$), is well below the overall time of diffusion. For fast diffusion processes surface concentration evolves in a significant way and the assumption of instantaneous equilibrium or saturation at the surface is no more justified. Analytical solutions to the time dependent first kind boundary condition are already available in the literature [7]. Here we present a new solution that presents advantages in computational approaches. The solution here presented has already been derived by the author [11] in the framework of interdiffusion processes occurring in ion exchange in silicate glass. A particular form of the time dependent boundary condition will be here justified on the basis of a surface saturation model. The physical-mathematical model is introduced based on the assumption of continuum media, a list of symbols with their description and units is reported in Table I.

## II. Concentration balance equation and constitutive relationship.

In the natural world there are quantities that are balanced through conservation mechanisms. Such quantities are in essence additive. The number of moles $M$ of a substance contained in a fixed domain $\Omega$ can be expressed in terms of its concentration $c(x,t)$ as follows:

$$M(t) = \int_{\Omega} c(x,t) dV .  \qquad (1)$$

The balance of the number of moles of a substance can be set by considering what causes $M$ to change with time. There can be identified two causes for change of $M$: internal sources or sinks "$\sigma$" distributed within the volume $\Omega$ and the exchange of moles with the external world, this last case can be represented by a total flux "$J_{tot}$". Figure 1 represents the fixed domain with its boundary where internal sources and boundary fluxes are indicated. The total quantity of moles contained in the volume of the domain changes in time because of the two identified causes: internal sources and total flux at boundaries.

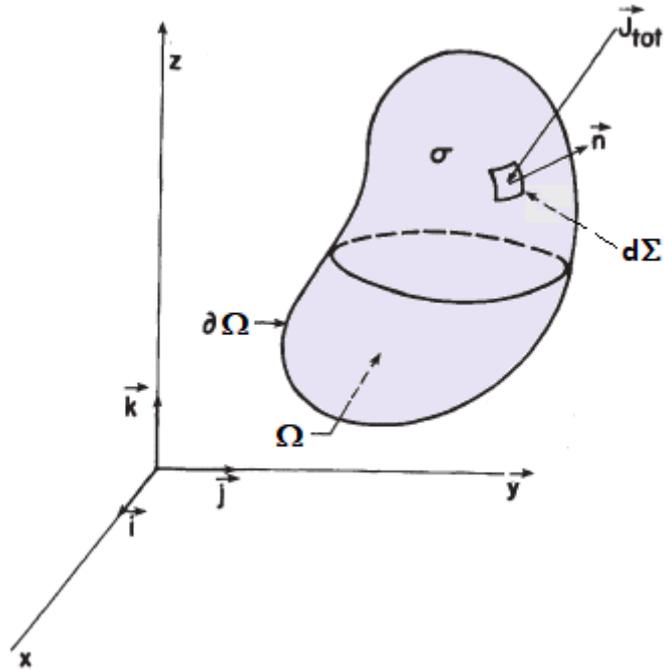

**Figure 1 – Fixed domain $\Omega$ with the internal source $\sigma$ and the total flux through the domain surface $\partial\Omega$. (figure 1 is largely taken from [10] with some minor changes).**

The balance equation is:

$$\dot{M}(t) = \int_\Omega \sigma dV - \oint_{\partial\Omega} \vec{J}_{tot} \cdot n d\Sigma \ . \tag{2}$$

From (1) and using the divergence theorem we can set a local balance equation:

$$\frac{\partial c}{\partial t} = -\vec{\nabla}\vec{J}_{tot} + \sigma \ . \tag{3}$$

The total flux is composed of a convective component due to the overall fluid motion with a velocity field $\vec{v}$ and a flux component $\vec{J}$ due to causes different from convection:

$$\vec{J}_{tot} = c\vec{v} + \vec{J} \quad , \tag{4}$$

with this position the balance equation is:

$$\frac{\partial c}{\partial t} + \vec{\nabla}(c\vec{v}) = -\vec{\nabla}\vec{J} + \sigma \ . \tag{5}$$

We have derived equation (5) by using the balance equation for a fixed volume $\Omega$, the same result can be obtained considering, in general, a domain of a continuous medium as a function of time $\Omega(t)$. Following this second approach and making use of the so-called Reynolds Transport Theorem (RTT) [12],[13] we obtain the same equation (5). The use of balance equation approach versus RTT approach is discussed in the literature [14]. Within the context of diffusion in solid bodies a preference towards balance equation can be envisaged as it results closer to a physical principle. Additionally, in solids, the convective velocity is zero and, when source term vanishes, the balance equation finally results:

$$\frac{\partial c}{\partial t} = -\vec{\nabla}\vec{J} \ . \tag{6}$$

Balance equation (6) is just a statement connecting causes (flux of matter) to effects (change in concentration). Solution to balance equation as written in (6) requires a relationship between concentration and flux. Such relationship is called "constitutive" as it comes from a microscopic model of mass transport process. From non-equilibrium thermodynamics, see Jost [1] page 156, Ansermet and Brechet [15], Flynn [16] and Mauro [17], a first order linear relationship can be set between flux and chemical potential through the diffusion coefficient:

$$-\vec{J} = \frac{cD}{RT}\vec{\nabla}\mu \ , \tag{7}$$

passing to one-dimensional formulation, equation (7) reads:

$$-J = \frac{cD}{RT}\frac{\partial \mu}{\partial x} \ . \tag{8}$$

In order to write an equation for the concentration, chemical potential is expressed in terms of concentration through the activity coefficient [1],[16]:

$$\mu(x,t) = \mu_0 + RT \ln[\gamma c(x,t)] \ . \tag{9}$$

The gradient of (9) is:

$$\frac{\partial \mu}{\partial x} = RT\left(\frac{\partial \ln \gamma}{\partial x} + \frac{\partial \ln c}{\partial x}\right) = RT\left(\frac{\partial \ln \gamma}{\partial \ln c}\frac{\partial \ln c}{\partial x} + \frac{\partial \ln c}{\partial x}\right) = \frac{RT}{c}\left(1 + \frac{\partial \ln \gamma}{\partial \ln c}\right)\frac{\partial c}{\partial x} \ , \tag{10}$$

with this position the constitutive relationship results:

$$-J = D\left(1 + \frac{\partial \ln \gamma}{\partial \ln c}\right)\frac{\partial c}{\partial x} = \tilde{D}\frac{\partial c}{\partial x} \, , \qquad (11)$$

Where it has been defined the chemical diffusion coefficient:

$$\tilde{D} = D\left(1 + \frac{\partial \ln \gamma}{\partial \ln c}\right) . \qquad (12)$$

Finally, after (11) and (6) we can set the mono-dimensional balance concentration equation

$$\frac{\partial c}{\partial t} = \frac{\partial}{\partial x}\left(\tilde{D}\frac{\partial c}{\partial x}\right) . \qquad (13)$$

Another additional assumption is to consider $\tilde{D}$ nearly constant and hence not depending on *x*. Based on this last assumption we obtain the well-known diffusion equation for concentration:

$$\frac{\partial c}{\partial t} = \tilde{D}\frac{\partial^2 c}{\partial x^2} \, . \qquad (14)$$

Equations (11) and (14) are also known respectively as first and second Fick equations.

### III.     Solutions to the diffusion equation: initial and boundary conditions.

The diffusion equation for concentration (14) is a second order linear partial differential equation (PDE). There can be found in the literature excellent monographic references treating both analytical and numerical solutions to this PDE [7],[8],[9],[10]. In general a solution to the diffusion equation in a domain exists and it is unique when initial and a boundary conditions are established. A popular solution in a semi-infinite domain is obtained by setting the initial condition:

$$c(x,0) = 0 \, , \qquad (15)$$

and a boundary condition:

$$c(0,t) = c_s \, . \qquad (16)$$

This means, in physical terms, that concentration at surface of the solid body is instantaneously achieved to a fixed constant value, $c_s$, and maintained at this level for all the diffusion process time. The solution to this problem (14), (15) and (16) is well known in the literature:

$$c(x,t) = c_s - c_s \text{erf}\left(\frac{x}{2\sqrt{\tilde{D}t}}\right) = c_s \text{erfc}\left(\frac{x}{2\sqrt{\tilde{D}t}}\right), \tag{17}$$

where the special function "$erfc(z)$" is the complementary error function:

$$\text{erfc}(z) = 1 - erf(z) = 1 - \frac{2}{\sqrt{\pi}} \int_0^z e^{-\xi^2} d\xi . \tag{18}$$

The assumption of an instantaneous constant surface concentration achievement is somehow critical [18]. In most of solid-state diffusion situations surface concentration is not achieved instantaneously and so maintained indefinitely at this equilibrium level. In general, boundary condition (16) is to be more realistically relaxed into a time dependent condition (boundary condition of first kind):

$$c(0,t) = g(t) . \tag{19}$$

Another type of boundary condition is considered by setting a balance condition on the flux of matter at the boundary, that means on the derivative of $c(x,t)$ (boundary condition of second kind,[9], [18]):

$$\left[\frac{\partial c(x,t)}{\partial x}\right]_{x=0} = \varphi(t) . \tag{20}$$

This boundary condition of the second kind will not be discussed here, an interesting discussion about boundary conditions of the second kind is introduced by Tagantsev [18] related to an interdiffusion problem generated by ion exchange in silicate glasses.

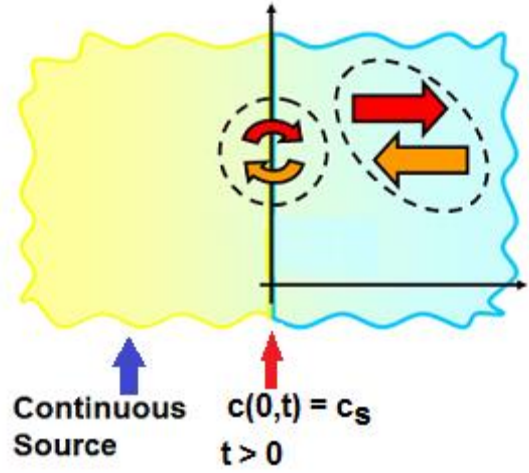

**Figure 2 – Semi-infinite solid with a surface boundary condition. Red arrows simple diffusion, Red and Orange arrows, case of interdiffusion.**

In this study we limit our attention to the following initial and boundary value problem (IBVP):

$$\frac{\partial c(x,t)}{\partial t} - D\frac{\partial^2 c(x,t)}{\partial x^2} = 0$$
$$c(x,0) = 0 \qquad (21)$$
$$c(0,t) = g(t)$$

Figure 2 represents the situation we have analyzed here. There can be identified two types of processes: a kinetic transport process in the bulk of the material and a surface dynamic equilibrium process. The two processes represent the diffusion in the bulk described by the balance equation (14) and the surface equilibrium represented by the boundary condition (16). This problem has classical solutions based on the Duhamel principle [7]:

$$c(x,t) = \frac{x}{2\sqrt{\pi D}} \int_0^t g(\tau) \frac{e^{-x^2/[4D(t-\tau)]}}{(t-\tau)^{3/2}} d\tau$$
$$c(x,t) = \frac{2}{\sqrt{\pi}} \int_{x/2\sqrt{Dt}}^{\infty} g(t - \frac{x^2}{4D\mu^2}) e^{-\mu^2} d\mu \qquad (22)$$

The equivalence between the two forms of the solution can be proved by a suitable change of variables. Taking the first line of Eq. (22) and performing a suitable change of variable:

$$\mu = \frac{x}{2\sqrt{D(t-\tau)}}; \tau = t - \frac{x^2}{4D\mu^2} \rightarrow \frac{d\mu}{d\tau} = \frac{x}{4\sqrt{D}(t-\tau)^{3/2}}$$ we arrive easily to the second form as:

$\frac{\sqrt{\pi}}{2} g(t - \frac{x^2}{4D\mu^2}) \frac{d}{d\mu} \text{erf}(\mu) = g(t - \frac{x^2}{4D\mu^2}) e^{-\mu^2}$. We can also write the second equation (22) in the following form:

$$c(x,t) = \int_{x/2\sqrt{Dt}}^{\infty} g(t - \frac{x^2}{4D\mu^2}) \frac{d}{d\mu} \text{erf}(\mu) d\mu , \qquad (23)$$

applying the partial integration theorem to Eq. (23) and a back change of variables, we can find an additional form of the solution to the IBVP (21):

$$c(x,t) = g(t) - g(0)\text{erf}(\frac{x}{2\sqrt{Dt}}) - \int_0^t \dot{g}(\tau) \text{erf}(\frac{x}{2\sqrt{D(t-\tau)}}) d\tau \qquad (24)$$

All three forms: eqs. (21), (22) and (24) represent the same solution, and the choice is made on the basis of computational convenience. Another form of the general solution to the IBVP (21) can be found from Eq. (24) with some algebraic manipulations:

$$c(x,t) = g(0)\text{erfc}\left(\frac{x}{2\sqrt{Dt}}\right) + \int_0^t \dot{g}(\tau) \text{erfc}\left(\frac{x}{2\sqrt{D(t-\tau)}}\right) d\tau . \qquad (25)$$

The analytical solutions (24) and (25) represent the incoming diffusing elements concentration, $c(x,t)$, in the solid after a defined diffusion time $t$ of contact of the solid with the diffusing elements.

The simplest choice of the $g(t)$ function is for a constant $g(t)=c_s$ at $t=0$ and kept constant for all $t>0$. This is exactly the case of immediate concentration equilibrium achieved between the solid and the diffusing elements source as they are put in contact (in simple batch process $t=0$ coincides with the time at which the solid is immersed in the diffusing elements source). In this case the derivative of $g(t)$ under the integral of (24) is zero while $g(0)=c_s$ hence:

$$c(x,t) = c_s - c_s \text{erf}\left(\frac{x}{2\sqrt{Dt}}\right) = c_s \text{erfc}\left(\frac{x}{2\sqrt{Dt}}\right) . \qquad (26)$$

As expected, solution (24) reduces to solution (17) when the boundary condition function $g(t)$ reduces to a constant.

## IV. Surface boundary condition: a solid-state model based on saturation of available sites.

The diffusing elements, entering in contact with the surface of a solid body, will occupy an available site. The occupation mechanisms will be substitutional, interstitial or through exchange of outgoing elements in interdiffusion processes. The first kind time dependent boundary condition (19) can be used to identify the availability on the surface, of sites to be occupied. Let's specify by $C_{rel}(x,t)$ the relative concentration. Indicating with $C_0$ the maximum value of the concentration in the solid body the relative concentration is:

$$C_{rel}(x,t) = \frac{c(x,t)}{C_o} \qquad . \tag{27}$$

Clearly $C_{rel}(x,t)$ is a limited function, $0 \leq C_{rel}(x,t) \leq 1$. The availability of sites at the solid body surface to be occupied is $(1-g(t))$. The time derivative of $g(t)$ is just the rate of saturation at the surface. We assume that the rate of sites saturation on the surface is proportional to the number of available sites. Based on this assumption and by considering a dimensional argument we can introduce a characteristic saturation time $\tau_{sat}$ to establish the following equation for $g(t)$:

$$\tau_{sat} \dot{g}(t) = 1 - g(t) \quad . \tag{28}$$

At $t=0$, $g(0)=0$ while, at $t \to \infty$, $g(t \to \infty)=1$. With these conditions the solution to the differential equation (28) is:

$$g(t) = 1 - e^{-t/\tau_{sat}} \qquad . \tag{29}$$

This function for the time dependent boundary condition represents a physical situation where the equilibrium at the interface between the source of diffusing elements and the solid body is not instantaneous but is progressively achieved following an exponential curve with a characteristic time $\tau_{sat}$. In Figure 3 a plot is reported for the boundary condition (29) with a characteristic saturation time of 3600s (1 hour). In [18] a physical model is introduced where the boundary condition is the result of an equilibrium condition between the flux of diffusing elements at the surface and the transfer of diffusing elements from the source to the surface. Because flux is expressed as first order spatial derivative of concentration it is evident that we obtain boundary condition of the second kind (20), where the function $\varphi(t)$ is expressed through a transfer (kinetic) coefficient and the difference between the concentration of diffusing elements in the source and on the solid surface. This model is not discussed here. In our discussion we will remain within the framework of the boundary condition of first kind expressed by equation (29).

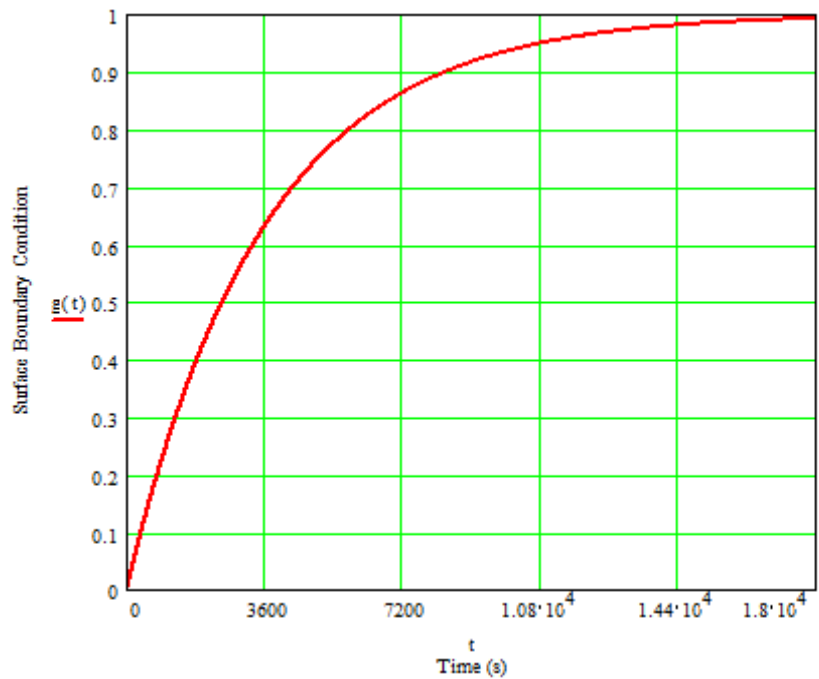

**Figure 3 – Surface boundary condition g(t) with a characteristic saturation time of 3600s (1 h).**

In this case $g(0)=0$ and $\dot{g}(t) = \dfrac{e^{-t/\tau_{sat}}}{\tau_{sat}}$, hence the solution of the IVBP (21) according to (24) is:

$$C_{rel}(x,t) = 1 - e^{-t/\tau_{sat}} - \int_0^t \frac{e^{-\tau/\tau_{sat}}}{\tau_{sat}} \operatorname{erf}\left(\frac{x}{2\sqrt{D(t-\tau)}}\right) d\tau \qquad (30)$$

In Figure 4 the relative concentration $C_{rel}(x,t)$ is calculated according to (30) considering a characteristic saturation time $\tau_{sat}=3600$s and total diffusion times of: one hour, $Crel1(x,t)$; two hours, $Crel2(x,t)$; three hours, $Crel3(x,t)$; four hours, $Crel4(x,t)$ and five hours $Crel5(x,t)$.

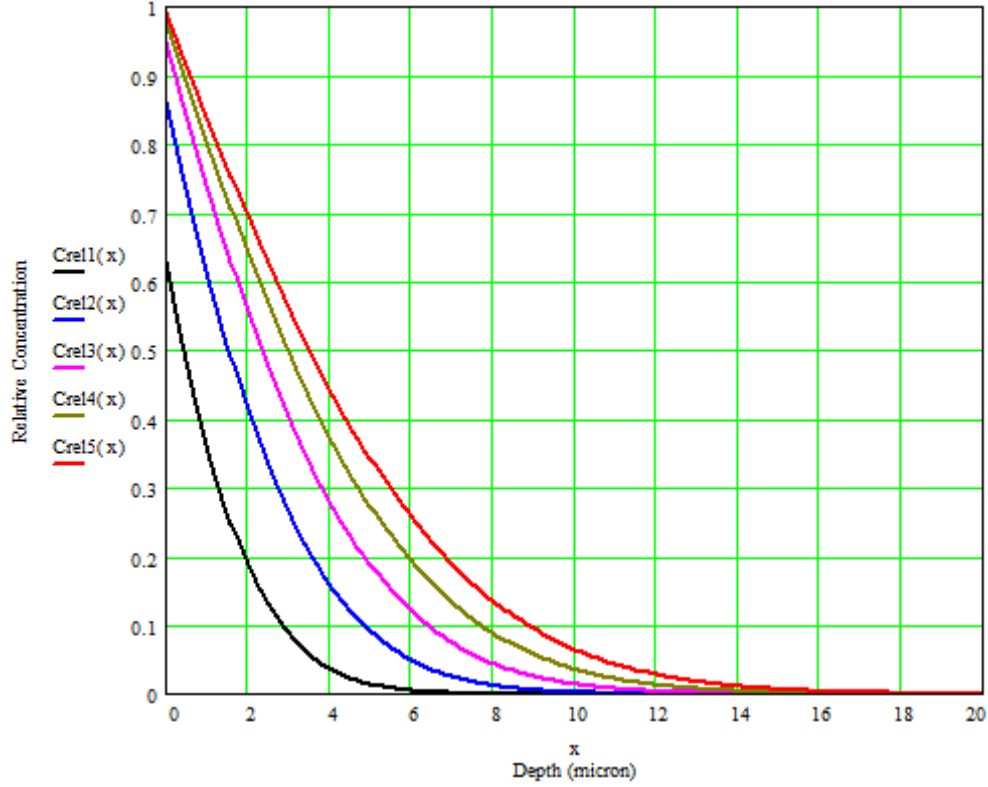

**Figure 4 – Relative concentration profiles calculated according to eqs. (30) for 4 different total diffusion tmes: *Crel1(x,t)* diffusion time 1hour, *Crel2(x,t)* diffusion time 2hours, *Crel3(x,t)* diffusion time 3hours, *Crel4(x,t)* diffusion time 4hours, *Crel5(x,t)* diffusion time 5hours. Saturation time $\tau_{sat}$=3600s (1h), diffusion coefficient *D*=1 10$^{-3}$ µm²/s.**

The characteristic penetration depth of the diffusing elements into the solid body can be roughly evaluated by the concept of diffusion length [2],[8]:

$$\lambda_D = 2\sqrt{D \cdot t} \qquad (31)$$

The values of relative surface concentration and diffusion length for the cases of Figure 4 are reported in Table II. As expected, the main result from the calculations reported in Figure 4 is that the effect of time dependent boundary condition is relevant when the characteristic saturation time at the solid surface is lower or at maximum 3 times the total diffusion time. As the total diffusion time become larger than 4 times the saturation time we can assume a constant value of the surface boundary condition. In this last situation the approximation of solution (30) with a simpler "*erfc*" behavior (26) is substantially acceptable.

## V.  Conclusion.

The main conclusions we can draw can be summarized as follows:

- A diffusion equation can be derived in the continuum media framework by setting a balance equation for mass transfer and a constitutive equation relating molar flux to its driving force namely the chemical potential.

- A time dependent first kind boundary condition at the solid surface is closer to a physical situation than a constant boundary condition which assume an instantaneous achievement of an equilibrium constant surface concentration.

- A new solution is presented equivalent to older ones [7] which is more convenient for computational purposes.

- A model is proposed based on the availability of surface sites to be saturated by the incoming diffusing elements leading to a new concept of characteristic saturation time $\tau_{sat}$.

- The PDE problem has been solved demonstrating that when the total diffusion time is larger than $4\tau_{sat}$ the solution can be conveniently approximated by the traditional "*erfc*" type solution.

**TABLES**

**Table 1 – List of symbols and units**

| Symbol | Description | Units |
|---|---|---|
| $M$ | Number of moles in a fixed volume $\Omega$ of a solid body. | mol |
| $x,y,z$ | Spatial coordinates | m |
| $t$ | Time | s |
| $c(x,t)$ | Concentration | mol/m$^3$ |
| $J_{tot}$, $J$ | Molar mass fluxes | mol/(m$^2$s) |
| $\sigma$ | Internal source | mol/(m$^3$s) |
| $v$ | Convective velocity | m/s |
| $D$ | Diffusion coefficient | μm$^2$/s |
| $T$ | Absolut Temperature | K |
| $R$ | Gas constant | J/(molK) |
| $m$ | Chemical potential | J/mol |
| $\gamma$ | Activity coefficient | Dimensionless |
| $\tau_{sat}$ | Characteristic saturation time | s |
| $\lambda_D$ | Diffusion length | μm |
| $C_{rel}$ | Relative concentration | Dimensionless |

**Table II – Surface concentration and diffusion length at different diffusion time. Characteristic saturation time $\tau_{sat}$= 3600s, Diffusion coefficient $D$=1 $10^{-3}$ µm²/s.**

| Total diffusion time – $t$ (s) | Relative surface concentration $Crel(0,t) = g(t)$ | Diffusion length - $\lambda_D$ (µm) |
|---|---|---|
| 3600 | 0.632 | 3.8 |
| 7200 | 0.864 | 5.4 |
| 10800 | 0.950 | 6.6 |
| 14400 | 0.981 | 7.6 |
| 18000 | 0.993 | 8.5 |